\documentclass[mypaper,7pt,twoside]{CoAst}
\usepackage{epsf,graphicx,fancyhdr}
\renewcommand{\chaptermark}[1]{\markboth{#1}{}}

\renewcommand{\headrulewidth}{0pt}
\newcommand{\kopf}{\small\itshape Comm. in Asteroseismology \\ Contribution to the Proceedings of the 38$^{th}$\,LIAC\,/\,HELAS-ESTA\,/\,BAG, 2008
}

\newcommand{\Authors}[1]{\begin{center}\normalsize\bf\sf #1 \end{center}}

\renewcommand{\author}[1]{\begin{center}\normalsize\bf\sf #1 \end{center}}
\newcommand{\Address}[1]{\begin{center}\small\sf #1 \end{center}}

\DeclareGraphicsExtensions{.eps,.jpg}

\newcommand{\Session}[1]{{\vspace{3mm}\small \noindent  \hspace*{3mm} Session: } #1 \normalsize}

\newcommand{\Objects}[1]{{\vspace{0mm}\small \noindent  \hspace*{3mm} Individual Objects: } \small #1 \normalsize}

	\newcommand{\three}{\small Atmospheres, mass loss and stellar winds}

\renewenvironment{abstract}{\section*{Abstract}\normalsize\sf}{}
\newcommand{\References}[1]{\begin{flushleft}{\large References\\}\vspace*{2mm}\small #1 \end{flushleft}}

\newcommand{\chapterCoAst}[2]{\chapter[\sf\normalsize #1\\ \footnotesize \hspace*{5mm}by #2 \sf\normalsize][]{#1\\}\rhead[\fancyplain{}{\sf\footnotesize \center{#1}}]{\fancyplain{}{\sffamily\thepage}}\lhead[\fancyplain{\kopf}{\sffamily\thepage}]{\fancyplain{\kopf}{\sf\footnotesize \center{#2}}}}

\newcommand{\figureCoAst}[5]{\begin{figure}[#4]
\centering
\includegraphics*[#5]{#1}
\caption{#2}
\label{#3}
\end{figure}}

\renewcommand{\chaptername}{}

\def\rfr{\smallskip\par\noindent
        \hangindent=7truemm
        \hangafter=1}

\begin{document}
\sf

\chapterCoAst{3-D Radiative Transfer Modelling of Massive-Star UV Wind Line Variability}
{A.\,Lobel} 
\Authors{A.\,Lobel$^{1}$} 
\Address{$^1$ Royal Observatory of Belgium \\
Ringlaan 3 - B 1180 Brussels - Belgium\\
}

\noindent
\begin{abstract}
We present detailed semi-empiric models for rotational modulations 
observed in ultraviolet wind lines of B0.5 supergiant HD\,64760. 
We model the Rotational Modulation Regions (RMRs) with advanced 3-D 
radiative transfer calculations in the stellar wind and quantitatively 
fit the time-evolution of the Si~{\sc iv} $\lambda$1395 resonance line. 
We find that the RMRs are due to linearly-shaped narrow sector-like density 
enhancements in the equatorial wind. Unlike the Co-rotating Interaction 
Regions (CIRs) which produce Discrete Absorption Components in the line, 
the RMRs do not spread out with larger distance above the stellar surface. 
The detailed best fit shows that the RMRs of HD\,64760 have maximum density 
enhancements of $\sim$17\% above the surrounding smooth wind density, 
about twice smaller than the hydrodynamic models of CIRs that warp 
around the star. The semi-empiric 3-D transfer modelling 
reveals that the narrow spoke-like RMRs must co-exist with  
broader and curved large-scale CIR wind density structures in the equatorial plane 
of this fast rotating Ib-supergiant.

\end{abstract}

\Session{ \three } 
\Objects{HD\,64760, $\xi$ Per, HD\,150168} 

\section*{Introduction}
Accurate mass-loss rates determined from quantitative spectroscopy  
are important to understand the wind-momentum luminosity relation of massive stars which
is affected by large-scale wind structures and clumping (Puls, 2008). 
Rotational modulations and Discrete Absorption Components (DACs) are important tracers 
of large-scale wind dynamics in massive hot stars. DACs are recurring absorption 
features observed in ultraviolet resonance lines of many OB-stars. 
They drift bluewards in the absorption portion of P-Cygni profiles and result from spiral-shaped 
density- and velocity-perturbations winding up in or above the plane of the equator 
(see Fig. 1 of Lobel 2008). Lobel \& Blomme (2008) demonstrated with 3-D radiative transfer (RT) 
modelling and hydrodynamic simulations for the detailed DAC evolution in HD\,64760 (B0.5 Ib) that 
these wind spirals are large-scale density waves caused by two unequally bright equatorial spots 
rotating five times slower than the stellar surface. Hydrodynamic models of structured winds with the 
density waves (for historical reasons they are termed 'Co-rotating Interaction Regions' or CIRs) 
reveal only a very small increase of less than 1\% in the smooth symmetric wind mass-loss rate.  
A characteristic property of the best fit CIR model is that the local increase of mass-loss 
from the bright spots creates lanes of enhanced density throughout the wind that always tend to broaden 
farther above the stellar surface. The CIR widths rapidly increase beyond their spot diameters, 
while the CIRs warp around the star over several tens of stellar radii, thereby producing two 
slowly migrating DACs with a period of 10.3 d. The rotational modulations of HD~64760, on the other hand, show 
a much shorter period of $\simeq$1.2 d and are morphologically very different from the DACs.
They are nearly-flat absorption components lasting only 0.5 d to 0.75 d with velocities that range
from $\sim$0 $\rm km\,s^{-1}$ to $\sim$$v_{\infty}$=1600 $\rm km\,s^{-1}$. They sometimes appear
to intersect the slower DACs and can reveal a remarkable 'banana' or bow-shaped 
intensity pattern with broad flux minima around $\sim$930 $\rm km\,s^{-1}$. 
Blomme (2008) showed that hydrodynamic spot models are unable to 
match the modulations of HD~64760 quantitatively (as opposed to the DACs) because 
$v_{\rm rot}$/$v_{\infty}$ $\simeq$0.16, with $v_{\rm rot}$ $\simeq$265 $\rm km\,s^{-1}$. 
The Rotational Modulation Regions (RMRs) in these spot models fail 
to reach $v_{\infty}$ sufficiently rapidly (e.g., as they turn too quickly 
around the star) and therefore lack the nearly linear absorption structure of the modulations above $\sim$1000 $\rm km\,s^{-1}$.

\figureCoAst{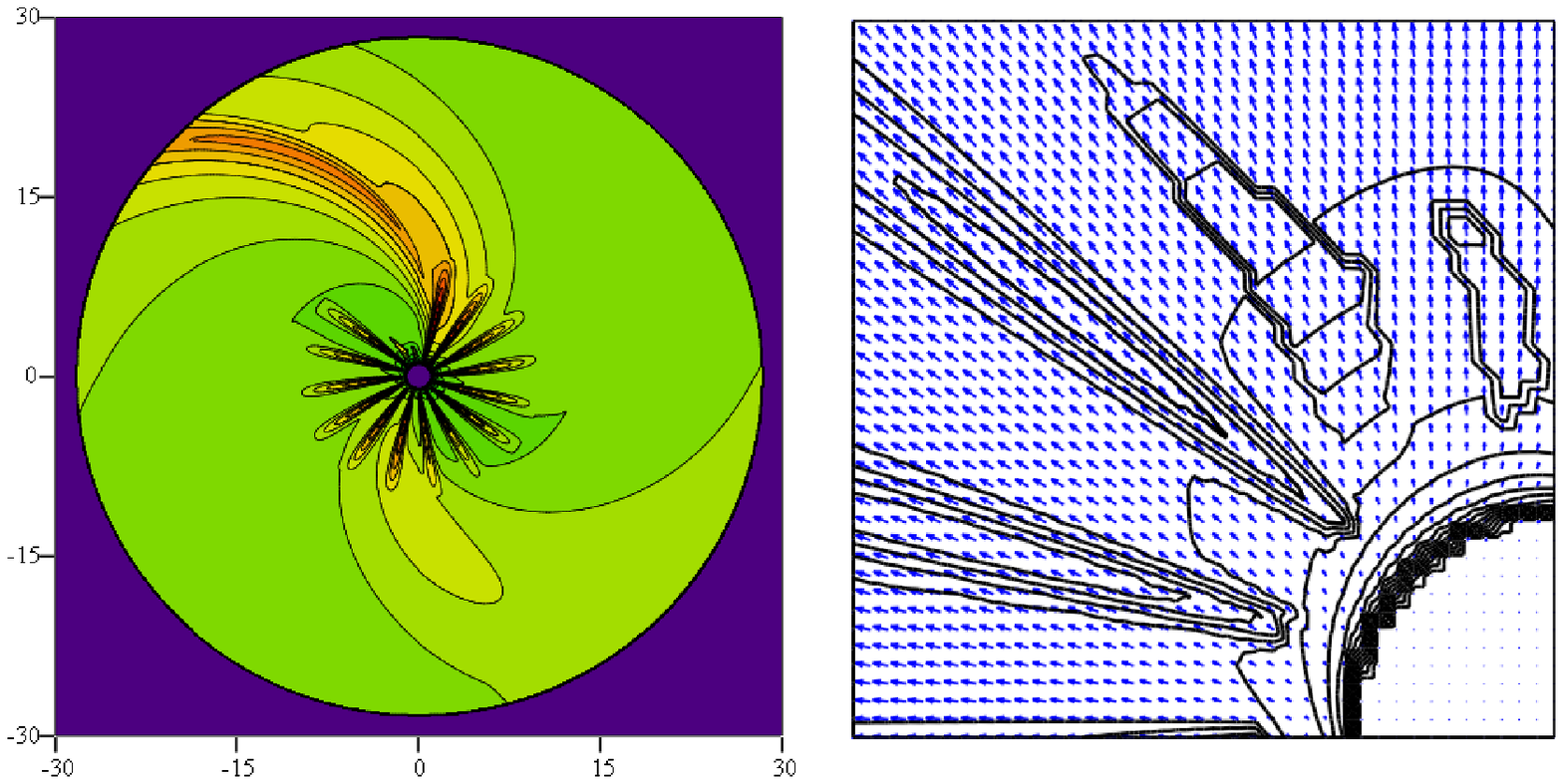}{Semi-empiric model for the equatorial wind density of HD~64760 
from 3-D radiative transfer fits to the dynamic spectrum of Si~{\sc iv} $\lambda$1395. 
The right-hand panel shows the bow-shaped density contours ({\it full drawn lines}) 
of the upper modulation in Fig. 3. Arrows mark the wind velocity in the hydrodynamic 
CIR model ({\it see text}).}{fig1}{!ht}{clip,angle=0,width=95mm}

In this paper we present a semi-empiric model for the RMRs that fits the modulations of
HD~64760 in detail. We adapt the hydrodynamic two-spot CIR model with parameterized wind 
density structures that quantitatively match the modulations. Our approach is 
motivated by constraining the geometry and density profiles of RMRs with 
3-D RT calculations using semi-empiric models before embarking upon 
more sophisticated multi-D hydrodynamic simulations.

\section*{3-D Radiative Transfer Modelling the Structured Wind of HD 64760}

We use the {\sc Wind3D} code developed in Lobel \& Blomme (2008) for detailed modelling 
the winds of stars with extended atmospheres. {\sc Wind3D} solves the non-LTE radiation 
transport problem in three geometric dimensions for arbitrary gas density structures 
and velocity fields, without assumptions of axial symmetry. The numerical transfer 
scheme is very accurate for tracing small variations of local density- and velocity-gradients 
on line opacity in strongly scattering-dominated supersonically expanding stellar winds.
The properties of the large-scale wind structures are constrained with detailed 3-D transfer 
fits to the slowly bluewards drifting DACs in Si~{\sc iv} $\lambda$1395 of HD\,64760. 
A hydrodynamic model with two spots of unequal brightness and size on opposite 
sides of the equator, 20\% and 8\% brighter than the stellar surface, and with 
opening angles of ${20}^\circ$ and ${30}^\circ$ diameter, provides the best 
fit to the DACs. 

While DACs are observed in many OB-stars, rotational modulations are often observed only 
in stars with large $v$sin$i$-values, such as HD~64760, $\xi$ Per (O7), and HD\,150168 (B1). 
Since these stars are observed almost equator-on it signals that RMRs are likely confined 
to the equatorial wind and become invisible toward smaller inclination angles. The RMRs 
therefore possibly result from unusual wind dynamics in the equatorial plane of fast 
rotating massive hot stars. Kaufer et al. (2006) investigate variability observed in optical lines 
of HD 64760 and propose a model with perturbations (spots) resulting from the interference 
of nonradial pulsations at the base of the wind.
     
\figureCoAst{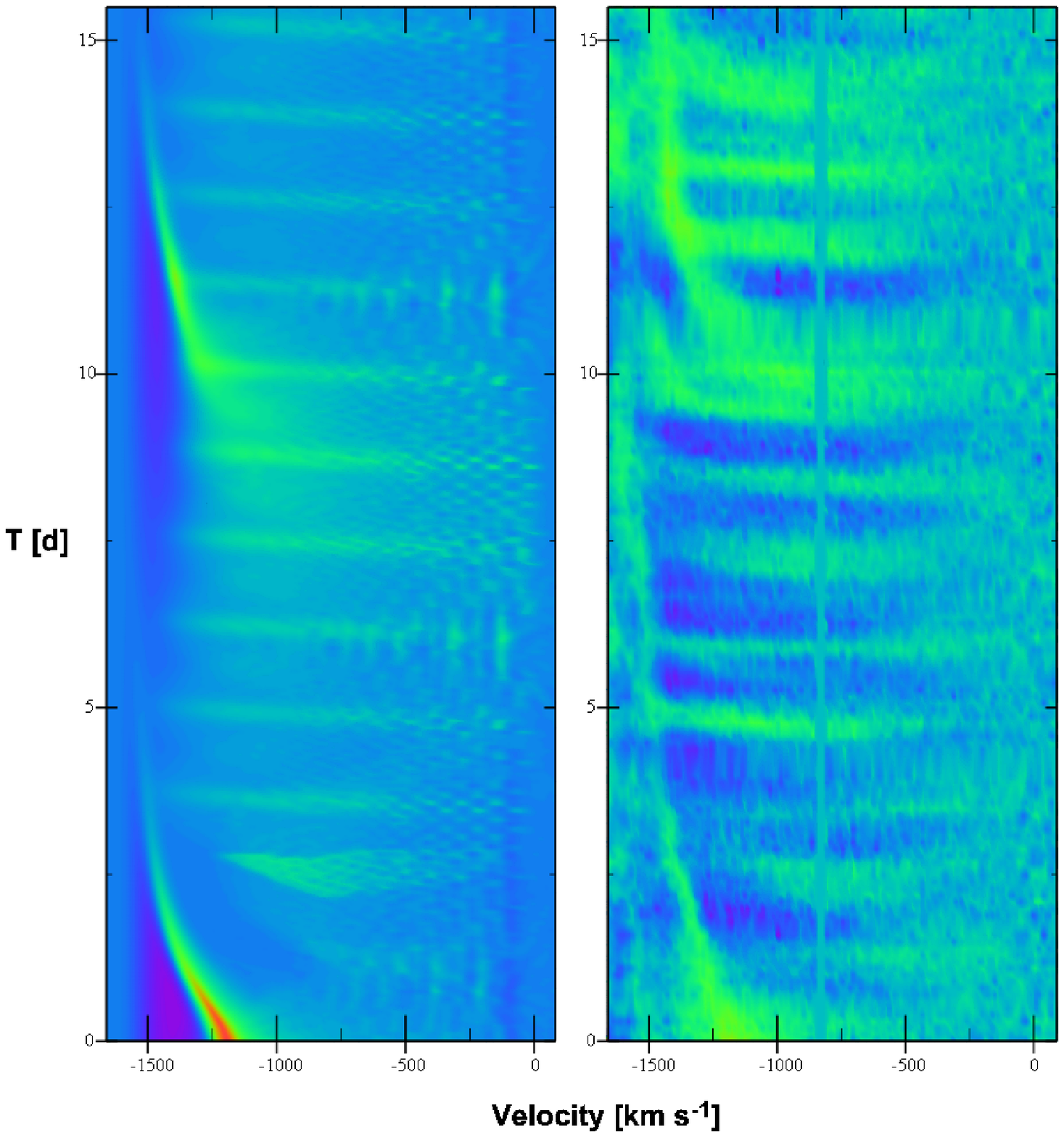}{Dynamic spectrum of Si~{\sc iv} $\lambda$1395 
observed during 15.5 d in HD\,64760 ({\it right-hand panel)} compared to 3-D radiative transfer
calculations ({\it left-hand panel}) with the wind model of Fig. 1. The horizontal
absorptions are modulations.}{fig2}{!ht}{clip,angle=0,width=75mm}

\section*{Detailed Modelling Rotational Modulations with Wind3D}

Figure 1 shows the best-fit density model computed with {\sc Zeus3D} for two CIRs in the wind of HD\,64760. 
We insert 16 central spoke-like RMRs with small opening angles in the CIR model that best fit 
the horizontal rotational modulations in the dynamic IUE spectrum of Si~{\sc iv} (Fig. 2). 
Our semi-empiric wind model for the modulations differs quantitatively from the kinematic model 
of Owocki, Cranmer, \& Fullerton (1995) because the RMRs do not appreciably curve over 
$\sim$10 $\rm R_{*}$. The RMRs extend very linearly (in narrow sectors) throughout the wind, away from 
the stellar surface, with maximum widths below 1~$\rm R_{*}$.    

Figure 3 shows a portion of the dynamic spectrum of Fig. 2 observed 
between 0~d and 3.1 d ({\it upper right-hand panel}). The lower DAC 
in Fig. 2 slowly migrates bluewards in time from $\sim$1100 $\rm km\,s^{-1}$ 
to $\sim$1400 $\rm km \,s^{-1}$, while two rotational modulations appear 
around 1.2 d ({\it lower modulation}) and 2.5 d ({\it upper modulation}).
The upper modulation is observed over $\sim$0.7 d and reveals a peculiar 'banana'-like shape, 
whereas the lower one occurs over $\sim$0.5 d with a more irregular 
absorption pattern below $\sim$800 $\rm km\, s^{-1}$.

\figureCoAst{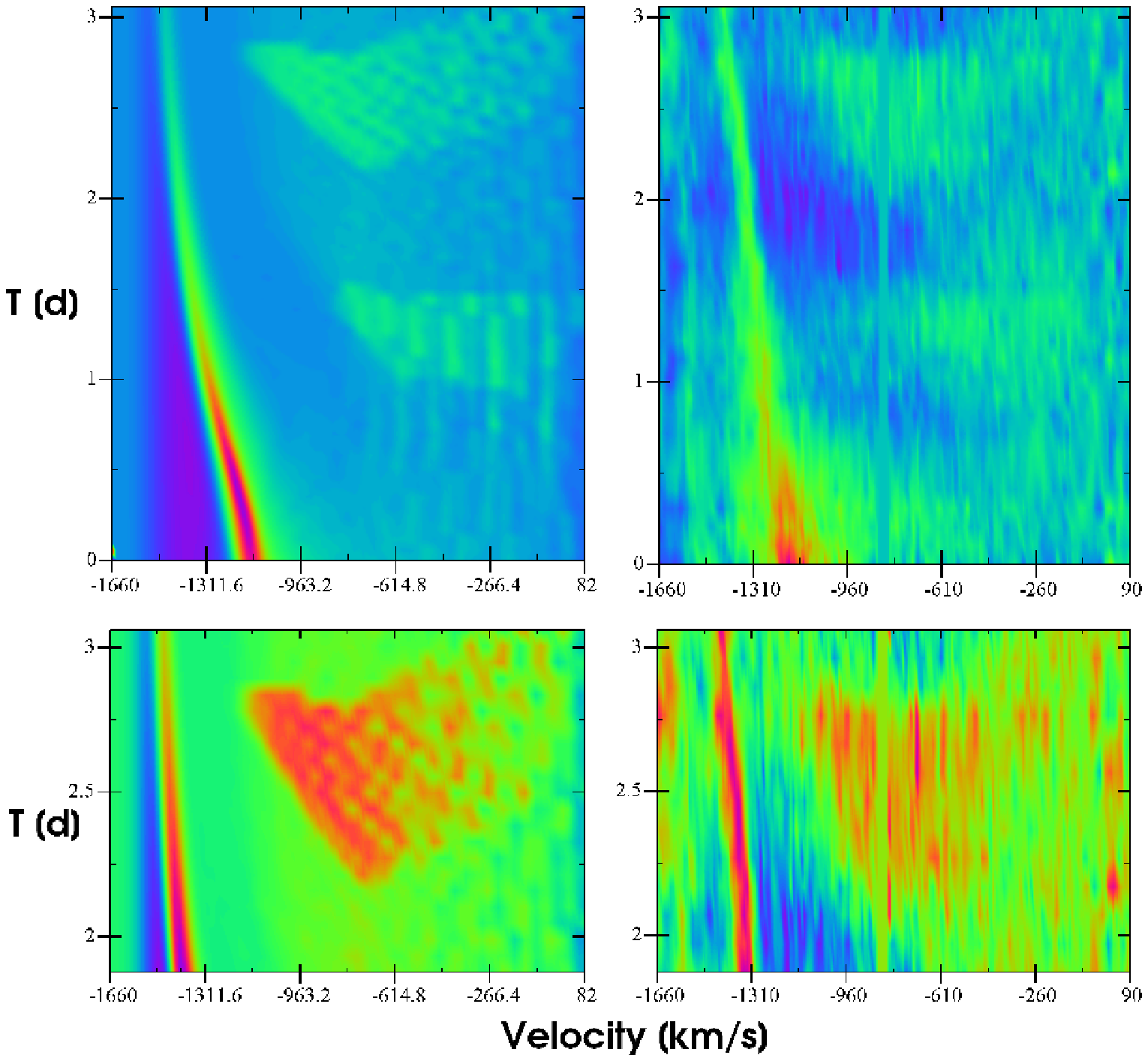}{Rotational modulations computed  with {\sc Wind3D} ({\it left-hand panels})
and observed ({\it right-hand panels}) in Si~{\sc iv} $\lambda$1395 of HD~64760. The lower panels show 
the best fit and observed upper 'bow-shaped' rotational modulation in more detail.}{fig3}{!ht}{clip,angle=0,width=85mm}

The upper left-hand panel of Fig. 3 shows the best fit with {\sc Wind3D}
using the wind density model of Fig. 1. 
We insert parameterized wind density structures in the CIR hydro model of HD\,64760
({\it left-hand panel of Fig. 1}) that quantitatively match the time- 
and velocity-position and relative absorption flux of the modulations. 
We semi-empirically delineate the borders of two RMRs (for the upper and lower modulations)
to compute the best-fit 
structured-wind density model. The lower panels of Fig. 3 show the best-fit 
theoretical ({\it left-hand panel}) and observed ({\it right-hand panel}) flux contrast 
of the upper modulation between 1.8~d and 3.1~d in the upper panels. 
The peculiar `wedge-like' shape at the short-wavelength side of the modulation 
is properly matched with the decrease of opening angle in the RMR density 
enhancement beyond 1~$\rm R_{*}$ above the stellar surface
({\it right-hand panel of Fig. 1}). The wind velocities in the upper modulation 
do not exceed $\sim$1200 $\rm km\,s^{-1}$ around 2.8~d. The upper modulation 
does not intersect the velocity position of the lower DAC around 1400 $\rm km\,s^{-1}$. 
The upper modulation density model in Fig.~1 does therefore not exceed a distance 
of $\sim$2.5 $\rm R_{*}$ above the stellar surface since the smooth wind velocity 
exceeds $\sim$1200 $\rm km\,s^{-1}$ only beyond that radius. We therefore 
attribute the remarkable 'banana'-shape observed in the upper modulation to 
the intrinsically bow-shaped front- and back-side density enhancement borders 
of the RMR in the right-hand panel of Fig. 1. 
The flux minimum observed in the upper modulation requires a RMR 
density maximum of $\sim$17\% above the smooth wind density around 
$\sim$3 $\rm R_{*}$ above the stellar surface in the model.   

The semi-empiric wind model of HD\,64760 in Fig. 1 provides a reliably 
accurate approximation of the equatorial wind structure we will investigate elsewhere with  
ab-initio 3-D hydrodynamic simulations. The narrow spoke-like RMRs are centered around 
the star with small inclination angles of $6^{\circ}$ from the radial direction, 
having equatorial opening angles of $10^{\circ}$. Similar as for both CIRs in the model, 
the RMRs are large-scale density- and velocity-structures in the fast radiatively 
accelerating equatorial wind. We do not semi-empirically model the RMRs beyond 10 
$\rm R_{*}$ above the surface because the modulations in Fig. 2 are observed up to $v_{\infty}$, 
while the wind velocity assumes already $\sim$95\% of $v_{\infty}$ at that distance. 
The 3-D RT modelling reveals that {\em linear} RMR wind density enhancements are required to 
quantitatively fit the horizontal modulations.
The RMRs are {\em linear} density structures in the equatorial wind because the 
modulations stay flat beyond 1000~$\rm km\,s^{-1}$.
The maximum RMR density contrast of $\sim$17\% is about half the maximum  
of $\sim$31\% in the CIR model of the lower DAC. The RMRs can therefore not appreciably alter 
the mass-loss rate of the smooth wind model. This was also concluded from detailed 
3-D RT modelling the DACs of HD\,64760 in Lobel \& Blomme (2008).

\section*{Conclusions}

We perform 3-D radiative transfer calculations for the rotational modulations in 
Si~{\sc iv} $\lambda$1395 of HD\,64760. We find with semi-empiric best fits that these 
horizontal line absorptions are caused by very linearly shaped density enhancements in 
the equatorial wind up to $\sim$10 $\rm R_{*}$. The RMRs do not exceed $\sim$17 \% of the 
smooth wind density, and do therefore not significantly change the stellar mass-loss rate. 
Lobel \& Blomme (2008) showed that the DACs in Si~{\sc iv} are caused by extended 
density waves of CIRs due to two bright equatorial spots that lag behind the surface 
rotation. We propose that the RMRs result from mechanical wave action producing narrow 
spoke-like wind collision regions around the star due to stellar 
nonradial pulsations at the fast-rotating wind base.

\References{
\rfr Blomme, R. 2008, in {\it Clumping in Hot-Star Winds}, 151, eds. W.-R. Hamann, 
A. Feldmeier, \& L. Oskinova, Universit\"{a}tsverlag Potsdam, Germany 
\rfr Kaufer, A., et al. 2006, A\&A, 447, 325
\rfr Lobel, A. 2008, in {\it Clumping in Hot-Star Winds}, 81, eds. W.-R. Hamann, 
A. Feldmeier, \& L. Oskinova, Universit\"{a}tsverlag Potsdam, Germany 
\rfr Lobel, A., \& Blomme, R. 2008, ApJ, 678, 408
\rfr Owocki, S., Cranmer, S., \& Fullerton, A. 1995, ApJ, 453, L37 
\rfr Puls, J., 2008, these proceedings
}

\end{document}